\documentclass[pdflatex,sn-mathphys-num]{sn-jnl}% Math and Physical Sciences Numbered  

\usepackage[utf8]{inputenc}  
\usepackage[T1]{fontenc}    
\usepackage{graphicx}      
\usepackage{xcolor}        
\usepackage{textcomp}       
 
\usepackage{amsmath,amssymb,amsfonts}  
\usepackage{amsthm}    
\usepackage{mathrsfs}       
\usepackage{stmaryrd}      
\usepackage{nicefrac}     

\usepackage{booktabs}       
\usepackage{multirow}       
\usepackage{tabularx}     

\usepackage{algorithm}      
\usepackage{algorithmicx}   
\usepackage{algpseudocode}   
\usepackage{listings}       
\usepackage{float}          
\usepackage{rotating}       
\usepackage{subcaption}     
 
\usepackage{manyfoot}       
\usepackage[title]{appendix} 

\usepackage{comment}        
\usepackage{microtype}     
\usepackage{pifont}        

\usepackage{hyperref}

\newcommand{\cmark}{\ding{51}}%
\newcommand{\xmark}{\ding{55}}%
\theoremstyle{thmstyleone}%
\theoremstyle{thmstyletwo}%

\theoremstyle{thmstylethree}%

\begin{document}

\title[Article Title]{DINO-LG: Enhancing Vision Transformers with Label Guidance for Coronary Artery Calcium Detection}


\author*[1]{\fnm{Mahmut S.} \sur{Gokmen}}\email{m.gokmen@uky.edu} 

\affil*[1]{University of Kentucky, Lexington, USA}

\author[1,2]{\fnm{Caner} \sur{Ozcan}}\email{canerozcan@karabuk.edu.tr} 
\affil*[2]{Department of Software Engineering, Karabuk University, Turkey} 

\author[1]{Moneera N. Haque}\email{mnha240@uky.edu} 
\author[1]{Steve W. Leung}\email{steve.leung@uky.edu} 
%\affiliation[4]{Department of Cardiology, Gill Heart and Vascular Institute, USA}

\author[1]{C. Seth Parker} \email{seth.parker@uky.edu}
 
\author[1]{W. Brent Seales} \email{seales@uky.edu}
\author[1]{Cody Bumgardner}\email{cody@uky.edu}


\abstract{Coronary artery disease (CAD), one of the leading causes of mortality worldwide, necessitates effective risk assessment strategies, with coronary artery calcium (CAC) scoring via computed tomography (CT) being a key method for prevention. Traditional methods, primarily based on UNET architectures implemented on pre-built models, face challenges like the scarcity of annotated CT scans containing CAC and imbalanced datasets, leading to reduced performance in segmentation and scoring tasks. In this study, we address these limitations by introducing DINO-LG, a novel label-guided extension of DINO (self-distillation with no labels) that incorporates targeted augmentation on annotated calcified regions during self-supervised pre-training. Our three-stage pipeline integrates Vision Transformer (ViT-Base/8) feature extraction via DINO-LG trained on 914 CT scans comprising 700 gated and 214 non-gated acquisitions, linear classification to identify calcified slices, and U-NET segmentation for CAC quantification and Agatston scoring. DINO-LG achieved 89\% sensitivity and 90\% specificity for detecting CAC-containing CT slices, compared to standard DINO's 79\% sensitivity and 77\% specificity, reducing false-negative and false-positive rates by 49\% and 57\% respectively. The integrated system achieves 90\% accuracy in CAC risk classification on 45 test patients, outperforming standalone U-NET segmentation (76\% accuracy) while processing only the relevant subset of CT slices. This targeted approach enhances CAC scoring accuracy by feeding the UNET model with relevant slices, improving diagnostic precision while lowering healthcare costs by minimizing unnecessary tests and treatments.}




\maketitle

\section{Introduction}

Cardiovascular disease (CVD) is the leading cause of death globally, responsible for approximately 17.9 million fatalities in 2019, which constitutes 32\% of all deaths worldwide \cite{CVD_2024}. Coronary artery disease (CAD), a major cardiovascular disease affecting the blood vessels that feed the heart muscle, caused 371,506 deaths in the United States in 2022 \cite{NCHS_2024}. According to the most recent heart disease and stroke statistics report, approximately 5\% of adults over the age of 20 have CAD in the United States \cite{Tsao2023-ae}. Early detection of CAD enables timely interventions that prevent disease progression and reduce heart attack risk, allowing for more effective treatments while lowering healthcare costs by avoiding serious complications.

Coronary artery calcium (CAC) scoring is considered a reliable tool for assessing cardiovascular disease and is generally recommended for use by various medical guidelines \cite{Knuuti2020-xr}. CAC scoring helps identify the presence and extent of calcified plaque in the coronary arteries which is strongly associated with the risk of CAD and future cardiovascular events. The test is non-invasive and relatively simple, using computer tomography (CT) scans to measure calcium deposits without the need for invasive procedures. The risk categorized calcium scores reflect different risk categories for cardiovascular events \cite{Oudkerk2008-zt}. Higher CAC scores can lead to more aggressive management of risk factor, while lower scores might support a more conservative approach. It is crucial for the radiographer to evaluate the position of high-density voxels in order to detect coronary calcification. CAC is typically measured using the Agatston method, assessing calcium deposits in the coronary arteries by measuring calcium density and volume to calculate a total calcium score \cite{Agatston1990-to}.  

The Agatston scoring system is a well-established method for quantifying CAC and stratifying cardiovascular risk \cite{Hecht2015-gd}. Evidence from large cohort studies, such as the Multi-Ethnic Study of Atherosclerosis (MESA), demonstrates a strong association between CAC scores and the risk of adverse cardiovascular events. For instance, in MESA, individuals with a CAC score of 0 were found to have a 10-year cardiovascular event risk of approximately 1\%, while those with scores >300 had risks exceeding 20\% over the same period. These findings underpin the use of CAC scoring in clinical guidelines, including those from the American College of Cardiology (ACC) and the American Heart Association (AHA), which recommend CAC assessment in intermediate-risk patients (10–20\% 10-year ASCVD risk) to refine risk stratification and guide therapeutic decisions. Notably, patients with scores $\geq$ 100 are often prioritized for statin therapy and other aggressive interventions, while those with a score of 0 may safely defer pharmacological treatments in favor of lifestyle modifications. Currently, the clinical analysis of calcium scores is performed semi-automatically by a software tool used by the radiologist to identify calcium regions by individually checking the slide images of each patient. This CAC measurement can be attention-demanding, labor-intensive, and time-consuming. To address these issues, automated CAC scoring methods have been developed, which can help enhance accuracy, consistency, and efficiency in measurements \cite{Eng2021-da,Van_Assen2021-mj,Takahashi2023-ah}.

Clinically, contrast-enhanced coronary CT angiography is a powerful imaging technique that uses contrast agent to provide detailed images of the coronary arteries to detect obstructive lesions and other vascular abnormalities, but involves higher radiation exposure due to the need for additional imaging sequences and the use of contrast \cite{Wolterink2016-vp}. On the other hand, non-contrast ECG-gated CT scans focus on quantifying coronary artery calcium, synchronizing image acquisition with the cardiac cycle to minimize motion artifacts and accurately assess calcified plaque for cardiovascular risk evaluation. Traditionally, when the CAC score has been calculated using non-contrast ECG-gated CT scans, a significant association with clinical outcomes has been observed \cite{Takx2015-ht,Chiles2015-eu}. However, other studies have shown that CAC calculation using non-contrast ECG non-gated CT data has excellent agreement with gated data \cite{Raygor2023-rv,Liu2022-vc,Zeleznik2021-ab,Kerndt2023-gy}. In this study, we propose a self-supervised learning based CAC scoring method that uses both gated and non-gated CT images together.

Most efforts in deep learning for image analysis have used some form of supervised learning. However, existing supervised learning methods face challenges when there is a lack of labeled data. In public datasets of CT scans, for instance, less than 10\% of the total slices contain calcified areas, despite the overall abundance of scans. This scarcity of labeled, relevant data poses a significant challenge for supervised approaches. Additionally, considering that calcifications typically cover only 3-5 millimeters, it becomes evident how challenging this task is. This small size makes accurate detection even more difficult, underscoring the need for advanced methods capable of identifying such minute details in CT scans.

Another learning paradigm that effectively addresses the mentioned data scarcity challenges is self-supervised learning (SSL) \cite{Nielsen2023-yx, Li2024-qq, Huang2023-mf}. Traditional supervised approaches face difficulties with the limited availability of labeled datasets, especially for calcified medical images. SSL allows models to learn from unlabeled data by extracting features without requiring explicit annotations \cite{Huang2023-mf}, enabling more robust training of algorithms to identify calcium deposits. This capability is particularly valuable for CAC scoring where labeled data is scarce. Despite successful automated CAC scoring methods using supervised learning, there is currently no reported method using SSL on both gated and non-gated non-contrast ECG CTs.

In this study, we have implemented one of the most popular SSL training technique using vision transformers (ViT) \cite{Xiao2023-wz} and commonly known as DINO (self-distillation with no labels) \cite{Caron2021-ci}. Regarding the architectural design of ViT models, generated features by ViT model in DINO model make possible to use these features in different tasks such as classification, segmentation or detection. In our approach, we employ Vision Transformer (ViT) models trained via the DINO technique to classify CT slices based on the presence of calcified areas. Given that calcifications typically occupy only 0.04\% of each CT slice and appear in less than 10\% of all slices in the dataset, generating features that effectively highlight and capture these small calcified regions presents a significant challenge.

To overcome this issue and generate the features capturing targeted areas' specifications,  we introduce a novel training method, DINO-LG, aimed at enhancing SSL approaches by incorporating label guidance to capture more specific features such as calcified areas in the model’s training process. This new technique is designed to contribute to SSL methodologies and expand upon existing training frameworks. Additionally, it has been demonstrated that vision foundational models can be directed toward specific areas, enabling them to generate features that effectively capture desired characteristics \cite{lg_1, lg_2}.

The remainder of this paper is organized as follows: Section 2 provides a review of related  work in automated CAC scoring and self-supervised learning. Section 3 presents our methodology, which includes the DINO-LG training approach and the classification and segmentation components. Section 4 describes the experimental setup, datasets, and quantitative results. Finally, Section 5 discusses our findings, limitations, and conclusions.

The main contributions of this paper are:
\begin{itemize}
\item \textbf{DINO-LG:} A novel label-guided training approach for self-supervised learning that directs the model’s attention to specific regions of interest through targeted augmentation, and is applicable beyond CAC detection to other medical imaging tasks with sparse annotations.
\item \textbf{Three-stage CAC scoring pipeline:} An integrated system combining DINO-LG feature extraction, linear classification for slice filtering, and U-NET segmentation, demonstrating improved accuracy (90\% vs. 76\%) while reducing computational burden through selective processing.
\item \textbf{Unified framework for gated and non-gated CTs:} A single model capable of processing both ECG-gated and non-gated CT scans, addressing a practical clinical need for versatile CAC assessment tools.
\end{itemize}
    
\section{Related Work} 
Recent reviews have highlighted notable advances in CAC scoring facilitated by AI techniques \cite{Aromiwura2024-by, Gennari2024-rs, Groen2024-mh, Abdelrahman2024-ea, Parsa2024-mx}. FDA-approved algorithms such as NANOX AI demonstrated 88.76\% agreement with cardiologists in analyzing 527 patients \cite{Kerndt2023-rm}. Sartoretti et al. \cite{Sartoretti2023-we} reported 0.986 similarity between CAC ground truth and DL model results using AVIEW CAC, while Assen et al. \cite{Van_Assen2021-mj} achieved a 0.921 correlation coefficient with AI-Rad Companion. These commercial systems have enabled rapid CAC assessment in routine clinical practice.

A novel DL model is developed by Eng et al. \cite{Eng2021-wv}  to automate CAC scoring, demonstrating high accuracy and speed on both dedicated gated coronary CTs and non-gated chest CTs. For detecting CAC, the algorithm provided a sensitivity of 71\% to 94\% and a positive predictive value of 88\% to 100\% across four different datasets. Velzen et al. \cite{Van_Velzen2020-hg} evaluated the effectiveness of a deep learning method for CAC scoring across different types of CT examinations to see if the algorithm result well when trained with representative images from different CT protocols. At baseline, the DL algorithm achieved intraclass correlation coefficients (ICCs) of 0.79-0.97 for CAC, which improved to 0.85-0.99 with combined training. Later, Zeleznik et al. \cite{Zeleznik2021-ki} introduced a deep learning system capable of accurately predicting cardiovascular events by quantifying coronary calcium, validated in a diverse population of 20,084 individuals on various routine cardiac-gated and non-gated CT. The study found a very high correlation of 0.92 (P < 0.0001) and substantial agreement between the automated and manual calculations of calcium risk groups. These studies highlight the increasing role of DL in improving clinical decision making in CAC detection.

Active multitask learning with uncertainty-weighted loss is proposed by Follmer et al. \cite{ Follmer2022-yw} for CAC scoring in ECG-gated CT.  The proposed model was tested on a total of 1,275 patient data consisting of a combination of three different datasets, and the results obtained showed a strong agreement with clinical outcomes ranging from 0.80 to 0.97. A 3D deep convolutional neural network (CNN) model is trained for Agatston scores using a database of 5973 non-contrast non-ECG gated chest CT without a prior segmentation of the CACs \cite{Gonzalez2018-cl}. This model achieved a Pearson correlation coefficient of r = 0.93; p $\leq$ 0.0001 when compared with ground truth data. In another study, Ihdayhid et al. \cite{ Ihdayhid2023-qp} proposed fully automated DL model to detect and measure CAC scores on ECG-gated CTs. The CAC score results from the automated model showed a strong correlation with the reference standard, indicated by a Spearman's correlation of r=0.90 and an ICC of 0.98. An automatic deep learning approach \cite{ Ihdayhid2023-qp} on 365 patients was developed to quantify CAC scores. The designed model was tested on an unseen cohort of 240 patients, and the results indicate that the model can effectively quantify CAC and classify risk in CT angiography. One of the studies using the public Coronary Calcium and chest CT's (COCA) dataset proposed a novel semantic-prompt scoring siamese network \cite{Li2023-vr}, while the other study proposed a lightweight 3D convolutions \cite{Santos2024-nk} with less memory requirements. The results obtained from these studies show that deep learning can produce similar results to the methods in the current literature in CAC scoring.

There have been studies using U-Net architecture \cite{Ronneberger2015-hk}, a convolutional neural network designed for efficient image segmentation, to automate the measurement of CAC from CT scans. This novel approach addresses the drawbacks of conventional manual scoring methods, which can be labor-intensive and require specialized knowledge. Gogin et al. \cite{Gogin2021-dn} evaluated 3D U-Net model to automatically estimate the amount of CAC on a database of 783 CT examinations and the final model resulted in a C-index of 0.951 on the test set. Bujny et al. \cite{Bujny2024-qz} introduced an algorithm for segmenting coronary arteries in multi-vendor ECG-gated non-contrast cardiac CT images, leveraging a new framework for semi-automatic Ground Truth (GT) creation through image registration. The model achieved a Dice coefficient of 0.65±0.08 for segmenting coronary arteries, evaluated against manually registered test GT. 

In another study \cite{Hong2022-yj}, the model was trained using focal loss, a modified version of cross entropy loss that emphasizes challenging samples by decreasing the weight of those that are easier to classify. A study of 1,811 retrospectively collected CT scans showed that U-Net++ significantly outperformed U-Net with accuracies ranging from 0.8 to 1.0. Since non-gated CTs are more commonly performed, \cite{Singh2022-ov} et al. developed a semi-supervised U-Net model to accurately assess risk in these scans, having been trained on gated scans from the before being applied to the non-gated images. The performance of the model showed a 91\% improvement in mean absolute error (with gated scans at 62.38 and non-gated scans at 674.19) and a 32\% increase in F1-score (gated scans at 0.68 compared to non-gated scans at 0.58). By effectively identifying and segmenting calcium deposition areas, the U-Net model improves both the speed and accuracy of CAC scoring, making it a useful tool for assessing cardiovascular risk. Its versatility across different imaging protocols and strong alignment with expert evaluations demonstrate its potential for broad clinical use, ultimately leading to earlier interventions and better patient outcomes in cardiovascular health.

Recently, in addition to traditional supervised learning methods, self-supervised learning (SSL) techniques use tasks that replace traditional labels with features extracted from the input data, allowing the model to learn data representations without requiring explicit supervised labels \cite{Huang2023-eo, Caron2021-ci, Oquab2023-vb}. This approach is particularly valuable for large-scale medical data, where labeling image data is often costly and time consuming. Shakouri et al. \cite{Shakouri2023-xt} highlighted the effectiveness of SSL, DINO, based on a vision transformer in enhancing chest X-ray classification. A quantitative analysis shows that the proposed method surpasses state-of-the-art techniques in accuracy and achieves similar performance in terms of AUC and F1-score, while requiring significantly less labeled data. Matsoukas et al. \cite{Matsoukas2023-by} demonstrated the advantages of SSL dealing with different scenarios. Through experimental studies, they showed that vision transformers can effectively replace CNNs for medical 2D image classification when appropriate training protocols are implemented. Additionally, Pérez-García et al. \cite{Perez-Garcia2024-xp} explored SSL and introduced RAD-DINO, a biomedical image encoder pre-trained solely on diverse biomedical imaging data. Baharoon et al. \cite{Baharoon2023-qv} developed foundational models for the medical field, using DINOv2 to benchmark disease classification and organ segmentation. Experimental studies have shown the potential to alleviate the problem of data annotation while improving model generalizability and robustness. Huang et al. \cite{Huang2024-nt} performed a a glioma grading task and comprehensive analysis of DINOv2 and ImageNet pre-trained models using three clinical modalities of brain MRI data. DINOv2 outperformed other models, especially when taking advantage of the frozen mechanism.

\section{Methodology}
\subsection{Overall Architecture}
The proposed architecture for automated coronary artery calcium (CAC) scoring consists of three main sequential steps:
\begin{enumerate}
    \item \textbf{Feature Extraction:} CT scan slices are processed through a Vision Transformer (ViT) backbone trained with DINO-LG.
    \item \textbf{Slice Classification:} A linear classifier identifies CT slices containing calcified areas.
    \item \textbf{Segmentation and Scoring:} A U-Net model segments calcified regions and calculates CAC scores.
\end{enumerate}

\begin{figure*}[htbp]
\centering
\includegraphics[width=1\linewidth]{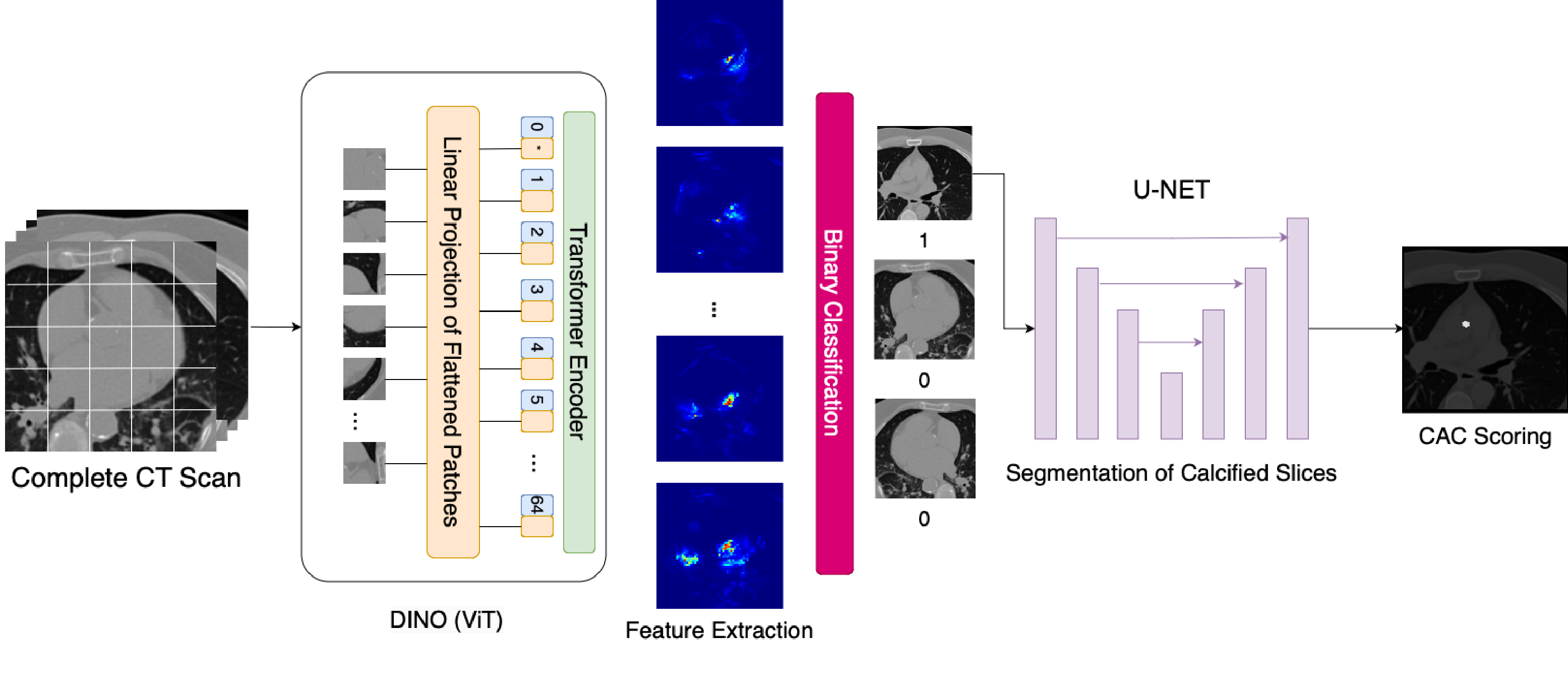}
\caption{Overall architecture of the proposed three-stage CAC scoring system. A complete CT scan is divided into slices (512$\times$512 pixels), which are processed through the DINO-LG backbone to extract 768-dimensional feature vectors. These features are fed to a linear classifier that identifies slices containing calcifications. Positively classified slices are then processed by a U-NET segmentation model to delineate calcified regions and compute Agatston scores. The visualized features in the figure are representative examples showing the discriminative power of DINO-LG embeddings.}

\label{fig:overall_arch}
\end{figure*}

As illustrated in Fig. \ref{fig:overall_arch}, our workflow begins with a complete CT scan, which is divided into smaller patches for efficient processing. These patches are fed into the ViT backbone of the DINO model, where each patch undergoes linear projection to extract features. The embedded features generated by the DINO model incorporate distinguishable characteristics for each CT slice, effectively differentiating between those containing calcification and those without.

In the second step, the extracted features are analyzed by a transformer-based binary classification model, which filters the CT slices and identifies those likely to contain calcified areas. This critical filtering step ensures that only relevant slices are passed to the computationally intensive segmentation stage.

Finally, the CT slices with a high probability of calcification are processed by a pre-trained U-Net model, which precisely segments the calcified regions. These segmentations are then used to calculate the total CAC score for the case according to standard clinical protocols.

This integrated approach focuses segmentation efforts on slices identified as containing calcifications, reducing computational overhead while addressing the challenges of limited labeled data and sparse distribution of calcifications.

\subsection{DINO-LG: Self-Supervised Learning with Label Guidance}
\subsubsection{Data Augmentation for Medical Imaging}
Data augmentation techniques are key components for self-supervised learning that enhance model robustness and improve generalizable representations by identifying underlying features. While standard RGB image augmentation includes techniques like solarization and color jittering, these are not suitable for medical images such as CT scans. Instead, we utilize augmentation methods specifically tailored for medical images, as outlined in Table \ref{tab:data_aug}.
The specific augmentation parameters used in our experiments are as follows: Random horizontal flips are applied with a probability of 0.5. Random resized crops extract regions with scales defined by the crop configuration (0.05-0.4 for local crops, larger scales for global crops). Gaussian blur uses a kernel size of 3$\times$3. For medical image-specific augmentations, brightness and contrast are adjusted within $\pm$40\% of the original values via ColorJitter. Additionally, random sharpness adjustment with a factor of 1.5 is applied with 80\% probability to enhance edge features relevant to calcification detection.

\begin{table} 
\caption{Comparison of augmentation techniques utilized for RGB images and medical images}\label{tab:data_aug}
\centering
\begin{tabular}{lcc} 
\toprule
\textbf{Augmentation Tech.} & \textbf{RGB Images} & \textbf{Medical Images}   \\
\midrule 
Random Horizontal Flip & \cmark & \cmark     \\
Random Vertical Flip & \cmark & \cmark     \\
Random Crop & \cmark &  \cmark   \\
Random Resized Crop & \cmark & \cmark   \\
Gaussian Blur & \cmark & \cmark     \\
Solarization & \cmark & \xmark     \\ 
Color Jitter & \cmark & \xmark      \\ 
Brightness Changing & \xmark & \cmark     \\ 
Noise Addition & \xmark & \cmark     \\ 
\midrule 
\end{tabular}
\vspace{0.3cm}
\end{table}

\subsubsection{Label-Guided Training Approach}
While random data augmentation techniques effectively increase overall accuracy and address imbalanced datasets, they may not optimally focus on specific regions of interest in medical images. Standard self-supervised learning approaches tend to generate generalized representations, which can lead to decreased performance in specialized tasks like calcification detection.

To address this limitation, we introduce a novel Label-Guided Data Augmentation technique that directs the model's attention to relevant calcified areas during training. In traditional DINO training, random local cropping (with ratios between 0.05 and 0.4 of the original image) constitutes the majority of augmentations. Our approach maintains this random component while adding label-guided cropping that specifically targets annotated calcified regions.

\begin{figure*}[h]
\centering
\includegraphics[width=1\linewidth]{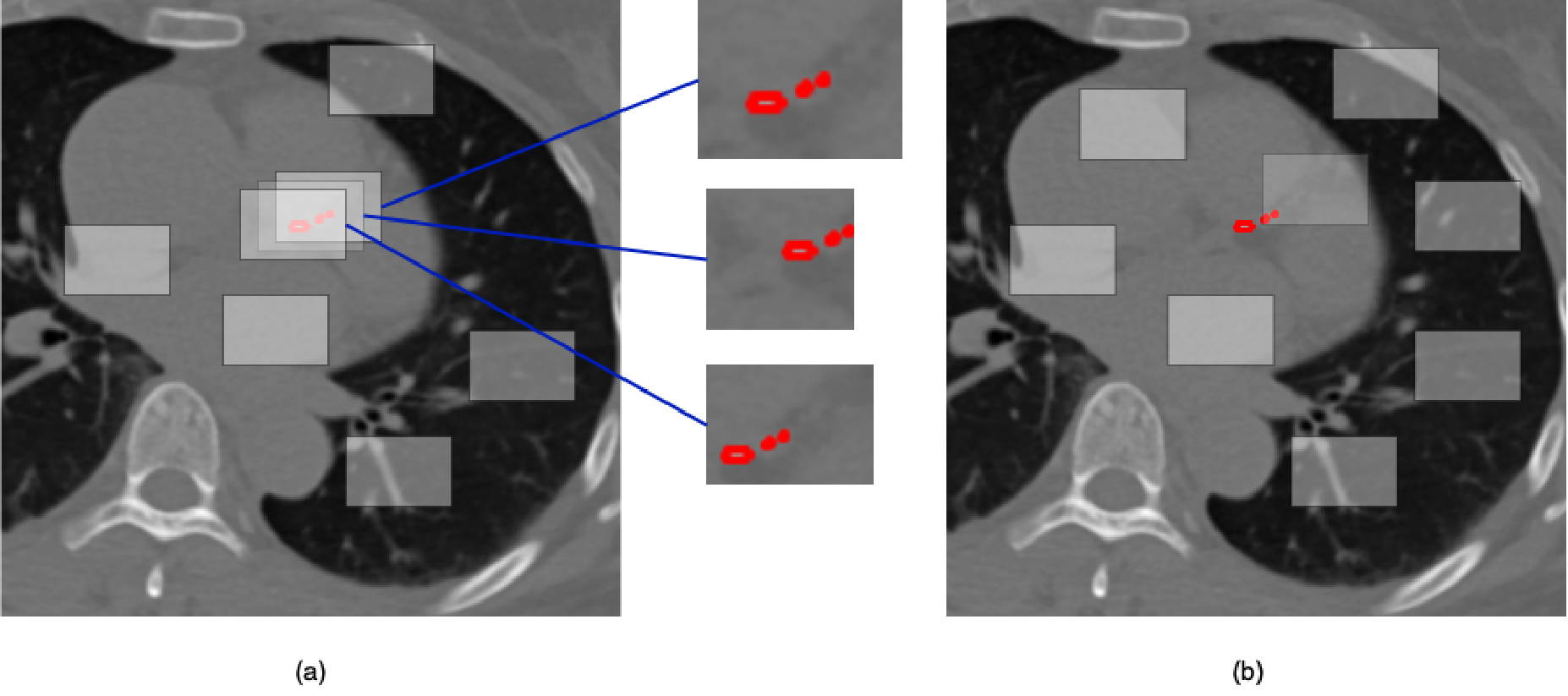}
\caption{Comparison of (a) DINO-LG and (b) standard DINO augmentation strategies.}
\label{fig:label_guided_illustration}
\end{figure*}

The label-guided process begins by selecting random coordinates from within the annotated calcified areas. These coordinates serve as focal points for generating crops that contain the regions of interest. This strategy is illustrated conceptually in Figure \ref{fig:label_guided_illustration}: DINO-LG generates both guided local crops centered on annotated calcified regions (shown by red bounding boxes) and random crops applied uniformly across the image, while standard DINO applies only random local crops without guidance. This dual-cropping approach enables the model to learn both calcification-specific features and general image representations. In our implementation, we generate 4 guided crops and 8 random crops per labeled image to balance specific feature learning with general representation capabilities.

The frequency of label-guided augmentations significantly impacts the model's focus on regions of interest. Too many label-guided augmentations may cause the model to neglect other useful features, while too few may not sufficiently highlight the calcified areas.

\begin{figure*}[h]
    \centering
    \begin{subfigure}[b]{0.48\textwidth}
        \centering
        \includegraphics[width=\textwidth]{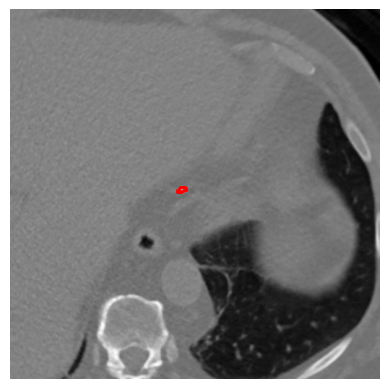}
        \caption{Annotated CT slice with annotation}
    \end{subfigure}
    \hfill
    \begin{subfigure}[b]{0.48\textwidth}
        \centering
        \includegraphics[width=\textwidth]{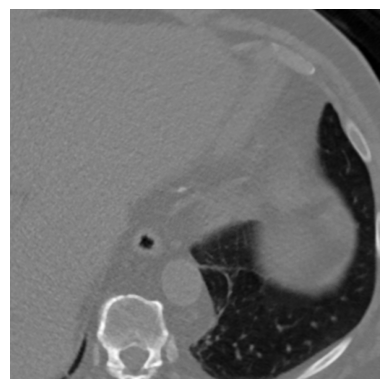}
        \caption{CT slice without annotation}
    \end{subfigure}
    
    \vspace{1em}
    
    \begin{subfigure}[b]{0.16\textwidth}
        \centering
        \includegraphics[width=\textwidth]{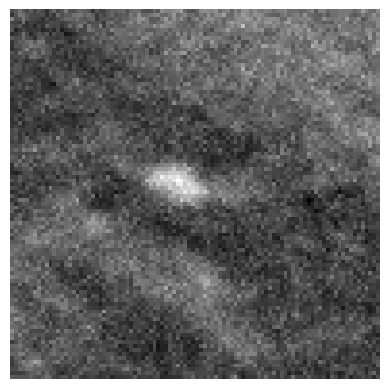}
        \caption{}
    \end{subfigure}
    \hfill
    \begin{subfigure}[b]{0.16\textwidth}
        \centering
        \includegraphics[width=\textwidth]{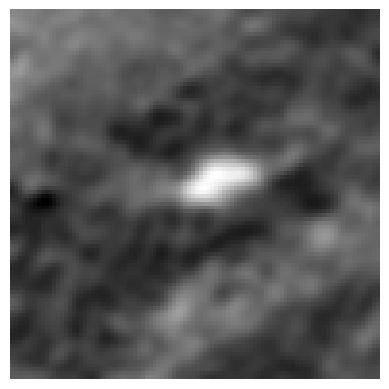}
        \caption{}
    \end{subfigure}
    \hfill
    \begin{subfigure}[b]{0.16\textwidth}
        \centering
        \includegraphics[width=\textwidth]{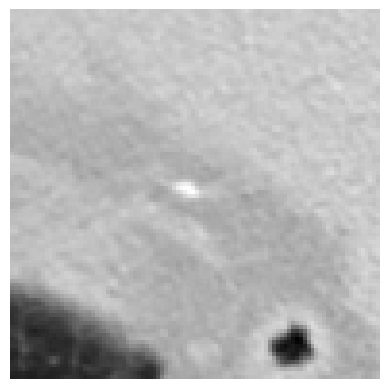}
        \caption{}
    \end{subfigure}
    \hfill
    \begin{subfigure}[b]{0.16\textwidth}
        \centering
        \includegraphics[width=\textwidth]{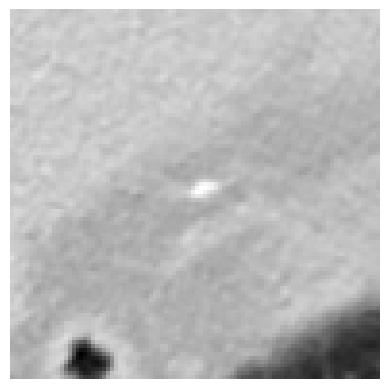}
        \caption{}
    \end{subfigure}
    \hfill
    \begin{subfigure}[b]{0.16\textwidth}
        \centering
        \includegraphics[width=\textwidth]{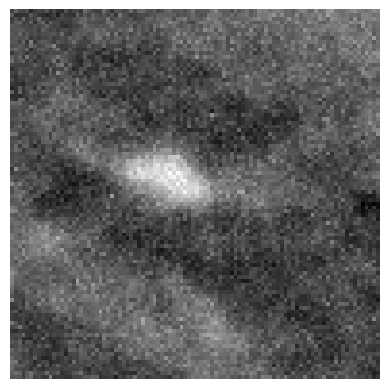}
        \caption{}
    \end{subfigure}
    \hfill
    \begin{subfigure}[b]{0.16\textwidth}
        \centering
        \includegraphics[width=\textwidth]{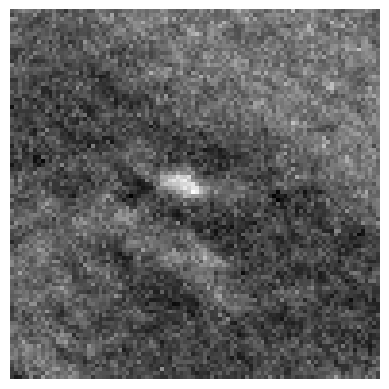}
        \caption{}
    \end{subfigure}
    \caption{Representation of guided local data augmentation. Figure $a$ represents labeled CT slice, while $b$ represents the same image without annotation. The figures from $c$ to $h$ show the guided local data augmentations utilized in DINO training.}
    \label{fig:guided_local_aug}
\end{figure*}

The frequency of label-guided augmentations significantly impacts the model's focus on regions of interest. In standard DINO training, 8 local crops are generated per image through random cropping. For our label-guided approach, when processing labeled slices (approximately 10\% of the dataset), we generate 4 additional guided local crops that specifically target annotated calcified regions. This results in a total of 12 local crops for labeled images (8 random + 4 guided) compared to 8 random crops for unlabeled images. The effective ratio of label-guided crops across the entire training dataset is approximately 5\% ($0.10 \times 4/12 \approx 0.033$ of all crops), which provides sufficient guidance without overfitting to labeled regions.
Figure \ref{fig:guided_local_aug} demonstrates our guided local data augmentation on a CT slice containing calcification, showing examples of the 4 guided local crops extracted from the annotated calcified regions.

\begin{figure*}[h]
    \centering
    \begin{subfigure}{0.24\textwidth}
        \centering
        \includegraphics[width=\linewidth]{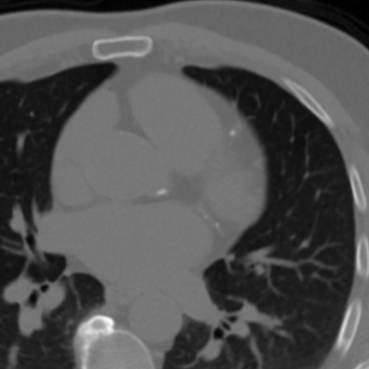}  
        \caption{}
        \label{fig:subimage1}
    \end{subfigure}
    \hfill
    \begin{subfigure}{0.24\textwidth}
        \centering 
        \includegraphics[width=\linewidth]{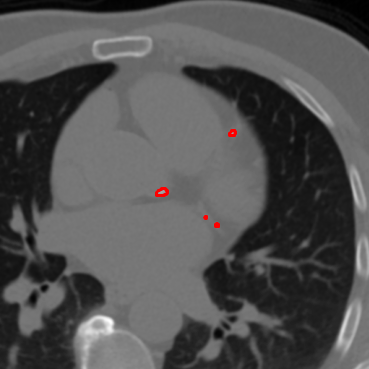}  
        \caption{}
        \label{fig:subimage2}
    \end{subfigure}
    \hfill
    \begin{subfigure}{0.24\textwidth}
        \centering
        \includegraphics[width=\linewidth]{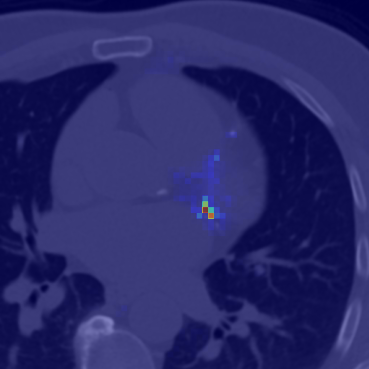}  
        \caption{}
        \label{fig:subimage4}
    \end{subfigure}
    \hfill
    \begin{subfigure}{0.24\textwidth}
        \centering
        \includegraphics[width=\linewidth]{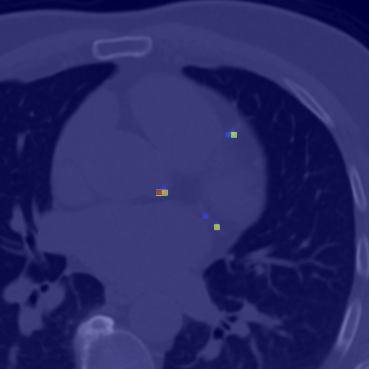}  
        \caption{}
        \label{fig:subimage3}
    \end{subfigure}
\caption{Feature visualization comparison via PCA. (a) Original CT slice. (b) Ground truth annotation. (c) Standard DINO features. (d) DINO-LG features.}
    \label{fig:attention_maps}
\end{figure*}

The effectiveness of our label-guided approach is visualized in Figure \ref{fig:attention_maps}, which compares feature representations generated by standard DINO and DINO-LG models via PCA analysis. The figure shows an original CT slice containing coronary calcification in the left anterior descending artery (a), where the ground truth annotation (b) occupies approximately 0.04\% of the image area. Standard DINO features (c) exhibit diffuse activation across cardiac structures, while DINO-LG features (d) demonstrate concentrated activation on calcified regions (bright yellow areas), validating successful feature guidance toward regions of clinical interest. The PCA visualization is applied to 768-dimensional [CLS] token embeddings extracted from the final transformer layer to illustrate how label-guided training directs the model's attention to task-relevant features.

\subsection{CT Slice Classification with DINO-LG Features}
The features generated by the DINO model are extracted from the last layer of multi-head self-attention modules. The length of these features equals the size of the embedded dimensions, which depends on the selected ViT model. In our approach, we focus on generating diverse features capable of capturing minute details such as calcified areas in CT slices.

For accurate classification of CT slices containing calcified areas, we implement a linear classification model trained on the features generated by the DINO model. This classifier processes the embedded representations to determine whether each slice contains calcification. The classification task is particularly challenging due to the small size of calcified areas relative to the overall CT slice and the imbalanced nature of the dataset, where only a small percentage of slices contain calcifications.

Our classifier architecture leverages the discriminative power of the DINO features to effectively distinguish between slices with and without calcification. The model is trained using a balanced loss function that accounts for the class imbalance, ensuring that both positive (calcified) and negative (non-calcified) slices are correctly classified. This is critical for downstream segmentation, as false negatives would result in missed calcifications while false positives would increase computational overhead without benefit.
The linear classifier consists of a single fully connected layer that maps the DINO feature vectors (768-dimensional for ViT-Base/8) to binary class probabilities. During training, the weights of the DINO-LG backbone remain frozen to preserve the learned self-supervised representations. We use the Adam optimizer with a learning rate of 1e-4, weight decay of 1e-4, and a batch size of 64. To address the significant class imbalance (approximately 10\% positive samples containing calcification), we use weighted binary cross-entropy loss with class weights inversely proportional to their frequencies in the training set. Specifically, the weight for the positive class (calcified) is set to 9.0, while the negative class (non-calcified) weight is 1.0. The model is trained for 10 epochs with early stopping monitored on the validation F1-score with a patience of 3 epochs. This limited training duration prevents the model from memorizing spurious patterns while ensuring convergence on the primary classification objective.

\subsection{Calcium Segmentation and Scoring}
After identifying CT slices likely to contain calcification through our binary classification model, we proceed to the segmentation stage. This step is crucial for precisely delineating the calcified regions and calculating accurate CAC scores.

For this purpose, we employ a U-NET architecture, which has demonstrated exceptional performance in medical image segmentation tasks. The U-NET model is specifically adapted for coronary calcium segmentation, with an encoder-decoder structure that effectively captures both local and global features. The encoder pathway extracts features through successive convolutional and pooling layers, while the decoder pathway reconstructs a detailed segmentation mask through up-sampling and skip connections.

By applying segmentation only to CT slices identified as containing calcification by our classifier, we significantly improve computational efficiency and reduce the potential for false positives. This targeted approach is particularly valuable given that calcified regions typically represent only a tiny fraction (approximately 0.04\%) of the total image area.

Once calcified regions are segmented, we quantify them using the Agatston scoring method, which considers both the density and volume of calcium deposits. This scoring system categorizes patients into four risk categories: low risk (0-10), moderate risk (11-100), high risk (101-400), and very high risk (>400). The accuracy of these risk assessments directly impacts clinical decision-making regarding preventive and therapeutic interventions.

The segmented regions are quantified using the Agatston scoring method, which incorporates both the area and density of calcium deposits. For each calcified lesion with a density threshold of at least 130 Hounsfield Units (HU) and minimum area of 1 mm$^2$, the Agatston score is calculated as:
$$\text{Agatston Score} = \sum_{i=1}^{n} A_i \times D_i$$
where $A_i$ is the area of the $i$-th calcified lesion in mm$^2$ and $D_i$ is a density weighting factor determined by the peak HU value of the lesion: $D_i = 1$ for 130-199 HU, $D_i = 2$ for 200-299 HU, $D_i = 3$ for 300-399 HU, and $D_i = 4$ for $\geq$400 HU. The cumulative score across all lesions categorizes patients into four risk categories as shown in Table \ref{tab:risk_category_counts}: low risk (0-10), moderate risk (11-100), high risk (101-400), and very high risk ($>$400).

\section{Experiments and Results}
\subsection{Experimental Setup}
The experimental setup consists of three steps of model training; starting with DINO model training, followed by the classification model, and finally training a U-NET for segmentation and CAC scoring. All experiments are performed on a single node with 8 H100 GPUs. 

\textbf{Dataset.} We utilized the COCA (Coronary Calcium and Chest CT) dataset from Stanford Hospital and Clinics \cite{COCA_Dataset}, comprising 1,003 CT scans in total: 789 gated and 214 non-gated acquisitions. The gated scans include pixel-level annotations for calcified regions in four coronary arteries: RCA (Right Coronary Artery), LAD (Left Anterior Descending), LCA (Left Coronary Artery), and LCX (Left Circumflex). Among all scans, 443 patients have at least one annotated slice, while 561 patients have no annotations. Each CT volume has dimensions of 512$\times$512 pixels with variable slice counts and a consistent 3 mm slice thickness. The annotated portion contains 36,411 slices, of which 3,656 (10\%) include calcifications, totaling 6,211 calcified objects with an average size of 119 pixels (approximately 0.04\% of each image area).

\textbf{Dataset Split and Usage.} The complete dataset composition and usage across different training stages is presented in Table \ref{tab:dataset_composition}. The full dataset comprises 1,003 CT scans: 789 gated (443 labeled, 346 unlabeled) and 214 non-gated (all unlabeled). Patients were stratified into training (914 patients), validation (44 patients), and test (45 patients) sets based on Agatston score distribution to ensure balanced representation across all risk categories (Table \ref{tab:risk_category_counts}). The training set comprises 700 gated scans (354 labeled + 346 unlabeled) and 214 non-gated scans (all unlabeled), totaling 914 patients. All 914 training patients were used for DINO-LG pre-training to leverage both labeled and unlabeled data through self-supervised learning. Subsequently, only the 354 labeled gated training patients were used to train the classification and segmentation models. The validation and test sets, containing exclusively labeled gated scans (44 and 45 patients respectively), were held out during all training phases and used solely for model evaluation.

\begin{table*}[h!]
\caption{Patient counts by risk category and Agatston score range for training, validation, and test sets}
\label{tab:risk_category_counts}
\centering

\resizebox{1.0\textwidth}{!}{%
\begin{tabular}{lcccc}
\toprule
\textbf{Risk Category} & \textbf{Range (Agatston Score)} & \textbf{Train Set} & \textbf{Validation Set} & \textbf{Test Set} \\
\midrule
Low Risk           & 0-10            & 45  & 5  & 6  \\
Moderate Risk      & 11-100          & 117 & 15 & 15 \\
High Risk          & 101-400         & 98  & 12 & 12 \\
Very High Risk     & $>$400          & 94  & 12 & 12 \\
\midrule
\textbf{Total Patients} &           & 354 & 44 & 45 \\
\bottomrule
\end{tabular}}
\vspace{0.3cm} 
\end{table*}
\begin{table*}[h!]
\caption{Complete dataset composition and usage across different model training stages. \cmark indicates the dataset was used in that training stage; \xmark indicates it was not used.}
\label{tab:dataset_composition}
\centering
\resizebox{1.0\textwidth}{!}{%
\begin{tabular}{lcccccc}
\toprule
\textbf{Dataset Split} & \textbf{CT Type} & \textbf{Labeled} & \textbf{Unlabeled} & \textbf{Total} & \textbf{DINO-LG} & \textbf{Class./Seg.} \\
\midrule
\multirow{4}{*}{Training} & Gated (labeled) & 354 & - & 354 & \cmark & \cmark \\
                          & Gated (unlabeled) & - & 346 & 346 & \cmark & \xmark \\
\cmidrule{2-7}
                          & \textit{Gated Subtotal} & \textit{354} & \textit{346} & \textbf{700} & & \\
\cmidrule{2-7}
                          & Non-gated (unlabeled) & - & 214 & 214 & \cmark & \xmark \\
\cmidrule{2-7}
                          & \textit{Training Total} & \textit{354} & \textit{560} & \textbf{914} & & \\
\midrule
Validation & Gated (labeled) & 44 & - & 44 & \xmark & \cmark \\
Test & Gated (labeled) & 45 & - & 45 & \xmark & \cmark \\
\midrule
\multicolumn{2}{l}{\textbf{Total Gated CT Scans}} & \textbf{443} & \textbf{346} & \textbf{789} & & \\
\multicolumn{2}{l}{\textbf{Total Non-gated CT Scans}} & - & \textbf{214} & \textbf{214} & & \\
\midrule
\multicolumn{2}{l}{\textbf{Grand Total}} & \textbf{443} & \textbf{560} & \textbf{1,003} & & \\
\bottomrule
\end{tabular}}
\vspace{0.3cm}
\end{table*}

\textbf{Data Preprocessing.} All CT scans underwent standardized preprocessing before model training. Each slice was first adjusted by subtracting the soft tissue reference value of 1024 HU to center the intensity distribution around zero. Subsequently, slice-level min-max normalization was applied to map HU values to the range [0, 1]:
$$I_{norm} = \frac{I - I_{min}}{I_{max} - I_{min} + \epsilon}$$
where $I$ represents the original intensity values, $I_{min}$ and $I_{max}$ are the minimum and maximum HU values within each slice, and $\epsilon = 10^{-7}$ prevents division by zero. Finally, intensities were linearly scaled to [-1, 1] via $I_{final} = 2 \times I_{norm} - 1$ to match the expected input range of the Vision Transformer architecture. This slice-wise normalization approach accounts for intensity variations across different CT acquisitions and scan protocols, ensuring consistent model inputs regardless of scanner manufacturer or acquisition parameters.

\textbf{DINO Training.} The original DINO repository provides sufficient guidance for training the DINO model on both single and multiple GPU setups. For DINO-LG pre-training, we utilize the complete training set of 914 patients comprising 700 gated CT scans (354 labeled + 346 unlabeled) and 214 unlabeled non-gated CT scans, as detailed in Table \ref{tab:dataset_composition}. This enables the model to learn robust self-supervised representations from both labeled and unlabeled data across different CT acquisition protocols. In our experiments, the same dataset is used to train two versions of the DINO model: the first using our label-guided training approach, and the second following the standard DINO training with data augmentations as described in Table \ref{tab:data_aug}. In addition to the hyper-parameters provided in the DINO repository such as the number of random local cropping (RLC), we have defined an additional hyper-parameter for our proposed method, referred to as the number of guided local cropping (GLC) given in Table \ref{tab:dino_train}.

\begin{figure*}[h]
\centering
\includegraphics[width=1.0\linewidth]{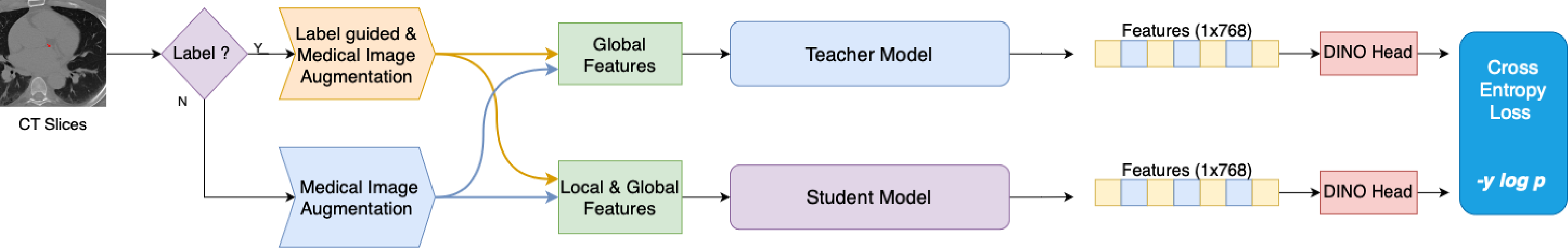}
\caption{The illustration of  training DINO-LG model.    }
\label{fig:dino_training}
\end{figure*}

A simplified representation of the label-guided training for DINO-LG model is shown in Figure \ref{fig:dino_training}. The main difference from traditional DINO model training is the integration of label checks for the corresponding CT slices. CT slices with annotations (from the 354 labeled gated scans) undergo our proposed label-guidance augmentation approach, while the remaining unlabeled CT slices (from 346 unlabeled gated and 214 non-gated scans) are augmented with the standard DINO model's hyper-parameters, with medically appropriate augmentation techniques as outlined in Table \ref{tab:data_aug}.

\begin{table*}[h]
\caption{Comparison of hyper-parameters for DINO-LG and standard DINO models. RLC: Random Local Crops per image (applied to all images). GLC: Guided Local Crops per labeled image (applied only to slices with annotations, in addition to RLC).}

\label{tab:dino_train}
\centering
\begin{tabular}{lccccc}
\toprule
\textbf{Model Name} & \textbf{ViT Type}   & \textbf{RLC}   & \textbf{GLC} & \textbf{Augmentation}   & \textbf{\# Epochs}   \\
\midrule 
DINO    & ViTb8    & 8 & 0   & Medical  & 150  \\ 
DINO-LG & ViTb8    & 8 & 4 & Medical & 150     \\
\midrule 
\end{tabular}
\vspace{0.3cm} 
\end{table*}
% Explain the experimental setup, including hardware, software, and the training procedure for both the DINO and UNET models.
 
% Detail the process of training the DINO model, including hyperparameters, epochs, and the optimizer used.

% Provide details on how the transformer-based classification model was trained and how the UNET model was trained for segmentation.

\textbf{Training Classifier.} The features generated by the DINO model are presented as input to the classification model to detect CT slices including calcified areas for each chest CT scans. It is also expected from the classification model to distinguish CT slices that are irrelevant to calcified areas. This is a crucial process to determine whether the features generated by the DINO model include calcified areas, even if these slices are not related to the chest or calcified areas.

\begin{figure*}[h]
\centering
\includegraphics[width=1.0\linewidth]{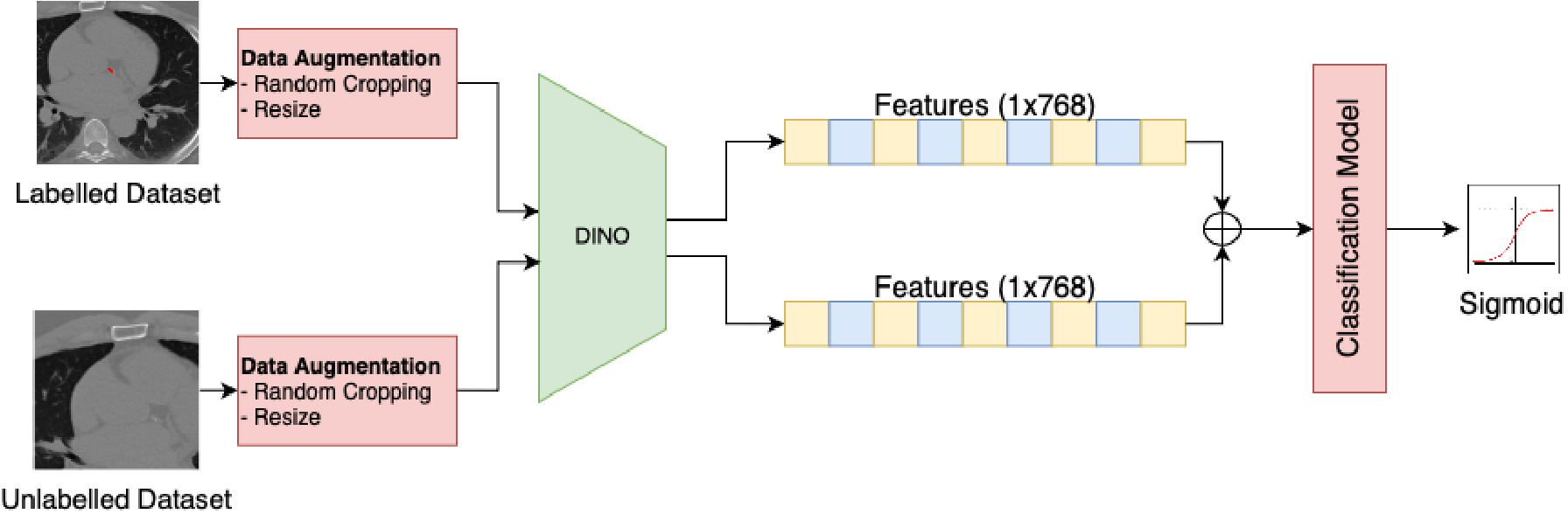}
\caption{The illustration of  training classifier model.    }
\label{fig:class_model}
\end{figure*}

The classifier is trained exclusively on the 354 labeled gated patients (Table \ref{tab:dataset_composition}), where approximately 10\% of slices contain calcifications, creating a natural class imbalance. Despite this imbalance, augmentation techniques such as random cropping and resizing are applied to the CT slices to enhance model robustness. Annotated slices serve as positive examples while non-annotated slices from the same patients serve as negative examples. The model is validated on 44 patients and tested on 45 held-out patients. The core architecture for training and implementing the classification model is illustrated in Figure \ref{fig:class_model}.

\textbf{Training Segmentation Model.} The segmentation models used in our experiments are based on U-NET-like architectures. Following the same data split as the classification task, the segmentation models are trained on 354 labeled gated patients, validated on 44 patients, and tested on 45 held-out patients (Table \ref{tab:dataset_composition}). The architectures we implemented include U-NET, Attention U-NET, and Swin U-NET, and their configurations are detailed in Table \ref{tab:segment_train}.

\begin{table*}[h]
\caption{Hyper-parameters for U-NET based segmentation models.}
\label{tab:segment_train}
\centering

\resizebox{1.0\textwidth}{!}{%
\begin{tabular}{lccccc}
\toprule
\textbf{Model Name} & \textbf{\# Initial Filter}   & \textbf{Image Size}   & \textbf{Batch Size} & \textbf{LR}   & \textbf{\# Epochs}   \\
\midrule  U-NET \cite{U_NET}   & 64    & 512x512 & 64 & 1e-4 & 100     \\
Attention U-NET \cite{attention_U_NET}    & 64    & 512x512 & 64 & 1e-4  & 100  \\ 
Swin U-NET \cite{swin_unet}   & 64    & 512x512 & 64 & 1e-4  & 100  \\ 
\midrule 
\end{tabular}}
\vspace{0.3cm} 
\end{table*}

\textbf{Segmentation Architecture Selection.} We conducted a comparative analysis of three U-NET-based architectures representing different evolutionary stages in medical image segmentation: (1) standard U-NET \cite{Ronneberger2015-hk}, the foundational convolutional architecture; (2) Attention U-NET \cite{attention_U_NET}, which incorporates attention gates for improved feature focus; and (3) Swin U-NET \cite{swin_unet}, a recent transformer-based approach utilizing hierarchical shifted window self-attention. Our empirical evaluation (Table \ref{tab:segment_train}) showed that standard U-NET achieved the best balance between segmentation performance (average Dice: 0.76 across all coronary arteries) and computational efficiency for our three-stage pipeline, making it the optimal choice for integration with DINO-LG feature extraction.

\subsection{Performance Analysis of DINO and Segmentation Models}

\textbf{Comparison of DINO-LG and DINO model via Classification}.
Classification performance is evaluated on the test set of 45 patients comprising 2,294 CT slices. Table~\ref{tab:comparison_classification} presents comprehensive performance metrics comparing standard DINO and DINO-LG models.

\begin{table*}[h!]
\caption{Comparison of Performance Metrics of Classification Task: Standard DINO vs. DINO-LG}
\centering
\begin{tabular}{lcc}
\hline
\textbf{Metric}                     & \textbf{Standard DINO} & \textbf{DINO-LG} \\ \hline
\vspace{0.03cm} \\ 
\multicolumn{3}{l}{\textbf{Confusion Matrix}} \\ \hline
True Negatives (TN)                 & 1491                  & \textbf{1744}             \\  
False Positives (FP)                & 446                   & \textbf{193}              \\  
False Negatives (FN)                & 76                    & \textbf{39}               \\  
True Positives (TP)                 & 281                   & \textbf{318}              \\   
\vspace{0.01cm} \\  
\multicolumn{3}{l}{\textbf{Classification Metrics}} \\ \hline
Class 0 Precision                   & 0.95                  & \textbf{0.98}             \\ 
Class 0 Recall                      & 0.77                  & \textbf{0.90}             \\  
Class 0 F1-score                    & 0.85                  & \textbf{0.94}             \\  
Class 1 Precision                   & 0.39                  & \textbf{0.62}             \\  
Class 1 Recall                      & 0.79                  & \textbf{0.89}             \\  
Class 1 F1-score                    & 0.52                  & \textbf{0.73}             \\  
Accuracy                            & 0.77                  & \textbf{0.90}             \\  
\vspace{0.01cm} \\ 
\multicolumn{3}{l}{\textbf{Sensitivity and Specificity}} \\ \hline
Sensitivity (True Positive Rate)    & 0.79                  & \textbf{0.89}             \\ 
Specificity (True Negative Rate)    & 0.77                  & \textbf{0.90}             \\  
\end{tabular}
\vspace{0.1cm}
\label{tab:comparison_classification}
\end{table*}
DINO-LG demonstrates substantial improvements across all metrics. Overall accuracy increases from 0.77 to 0.90, while sensitivity and specificity improve from 0.79 and 0.77 to 0.89 and 0.90, respectively. For individual classes, DINO-LG achieves higher F1-scores: 0.94 for Class 0 (non-calcified) and 0.73 for Class 1 (calcified), compared to 0.85 and 0.52 for standard DINO. Most critically, false-negative predictions decrease by 49\% (76→39) and false-positive predictions decrease by 57\% (446→193). McNemar's test confirms these improvements are statistically significant (p < 0.001).

\begin{figure*}[h!]
    \centering
    \begin{subfigure}{0.45\textwidth}
        \centering
        \includegraphics[width=\linewidth]{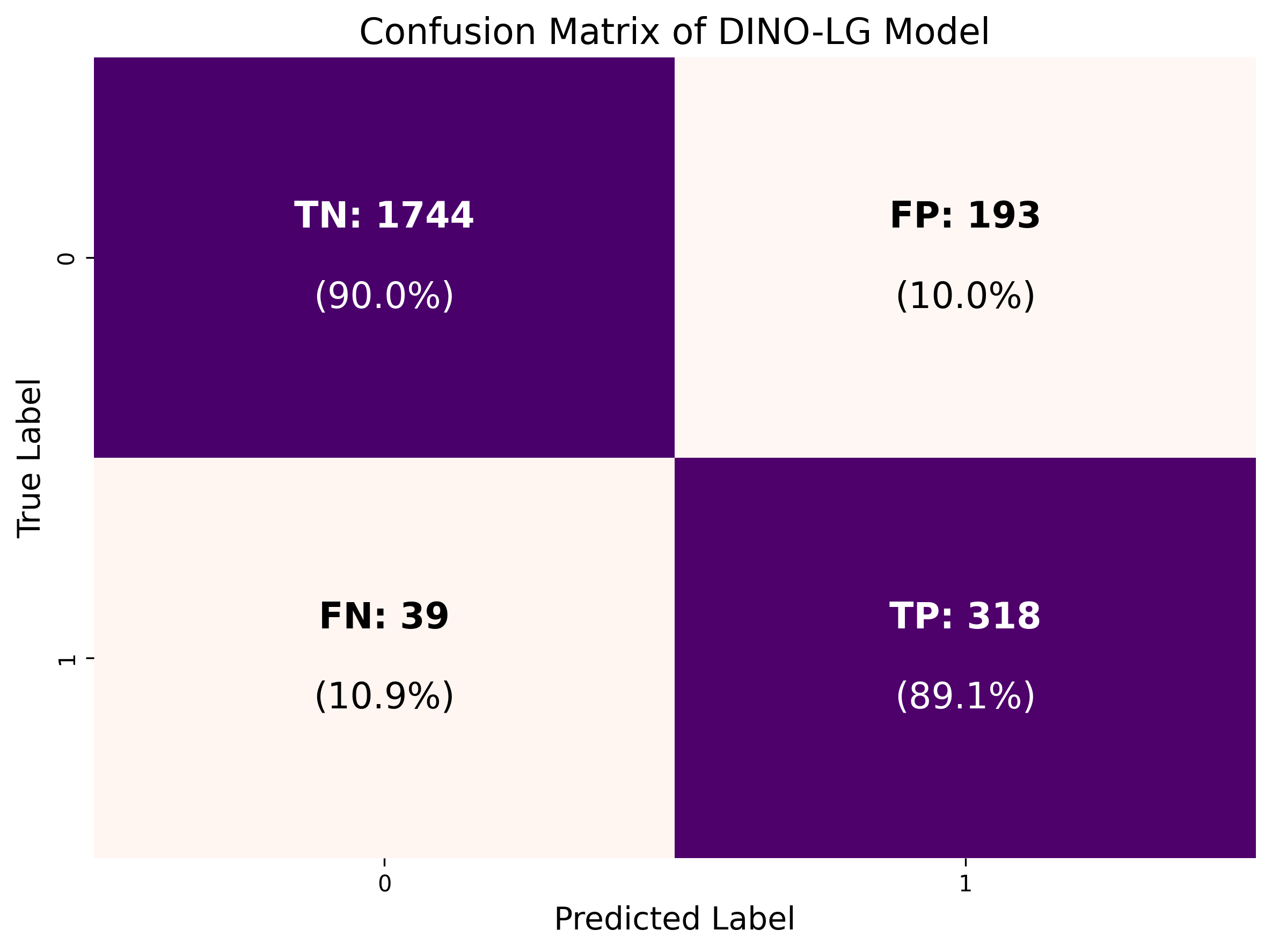}  
        \caption{DINO-LG classification model.}
        \label{fig:f1_score}
    \end{subfigure}
    \hfill
    \begin{subfigure}{0.45\textwidth}
        \centering
        \includegraphics[width=\linewidth]{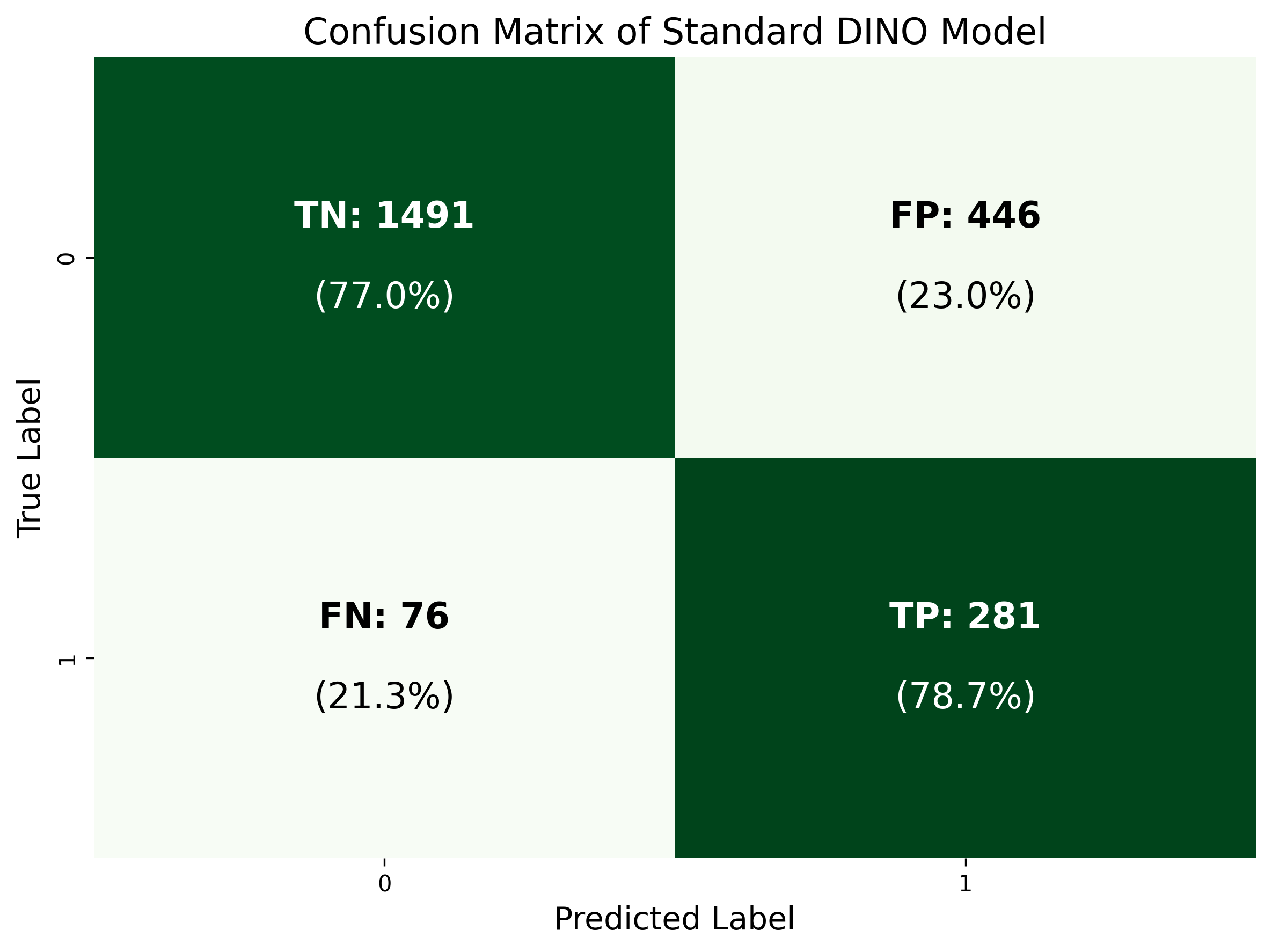}  
        \caption{Standard DINO classification model.}
        \label{fig:loss_graph}
    \end{subfigure}
\caption{Confusion matrices for (a) DINO-LG and (b) standard DINO on the test set of 2,294 CT slices from 45 patients.}
    \label{fig:classification_results}
\end{figure*}

The confusion matrices in Figure~\ref{fig:classification_results} illustrate these improvements visually. DINO-LG achieves balanced performance across both classes with 89\% sensitivity and 90\% specificity, while standard DINO shows notably higher error rates with 79\% sensitivity and 77\% specificity. The substantial reduction in false negatives (76→39) is clinically significant as it directly reduces missed calcifications.

The reduction in false-negative predictions is clinically critical, as these represent missed diagnoses where calcified slices are incorrectly classified as non-calcified. While false-positive predictions lead to unnecessary segmentation processing by classifying non-calcified slices as calcified, they are less concerning clinically as the subsequent U-NET segmentation stage can appropriately filter them.

\textbf{Segmentation Experiments}
The segmentation models were tested on the same dataset used in the classification experiments, independently of the classification model. Before integrating the segmentation model with DINO features, this section evaluates U-NET-based architectures to determine their suitability for combination with the DINO model in the main structure. The performance of U-NET-based models is presented in Table \ref{tab:segmentation_scores}. As shown in the table, the basic U-NET \cite{U_NET}, Swin U-NET \cite{swin_unet} and Attention U-NET \cite{attention_U_NET} architectures achieved significantly higher scores across all coronary artery classes: RCA (Right Coronary Artery), LAD (Left Anterior Descending), LCA (Left Coronary Artery), and LCX (Left Circumflex).

\begin{table*}[h!]
\caption{IoU and Dice scores for RCA, LAD, LCA, and LCX classes across three U-NET based segmentation models on the Test Dataset}
    \label{tab:segmentation_scores}
    \centering
    
\resizebox{1.0\textwidth}{!}{%
\begin{tabular}{lcccccccccc}
    \toprule
    \textbf{Model Name} & \multicolumn{2}{c}{\textbf{RCA}} & \multicolumn{2}{c}{\textbf{LAD}} & \multicolumn{2}{c}{\textbf{LCA}} & \multicolumn{2}{c}{\textbf{LCX}} & \multicolumn{2}{c}{\textbf{Average}} \\
    \cmidrule(lr){2-3} \cmidrule(lr){4-5} \cmidrule(lr){6-7} \cmidrule(lr){8-9} \cmidrule(lr){10-11}
                       & \textbf{IoU} & \textbf{Dice} & \textbf{IoU} & \textbf{Dice} & \textbf{IoU} & \textbf{Dice} & \textbf{IoU} & \textbf{Dice} & \textbf{IoU} & \textbf{Dice} \\
    \midrule
    Attention U-NET \cite{attention_U_NET}    & 0.95 & 0.92 & \textbf{0.63} & \textbf{0.77} & 0.36 & 0.53 & \textbf{0.59} & \textbf{0.74} & 0.63 & 0.74 \\
    Swin U-NET \cite{swin_unet}        & 0.96 & 0.96 & 0.52 & 0.68 & 0.10 & 0.16 & 0.35 & 0.52 & 0.48 & 0.58 \\
    \textbf{U-NET} \cite{U_NET}              & \textbf{0.97} & \textbf{0.98} & 0.61 & 0.75 & \textbf{0.43} & \textbf{0.60} & 0.49 & 0.70 & \textbf{0.63} & \textbf{0.76} \\
    \bottomrule
\end{tabular}
}
    \vspace{0.3cm}
\end{table*}
\hfill

\begin{figure*}[h!]
    \centering
    \begin{subfigure}{0.45\textwidth}
        \centering
        \includegraphics[width=\linewidth]{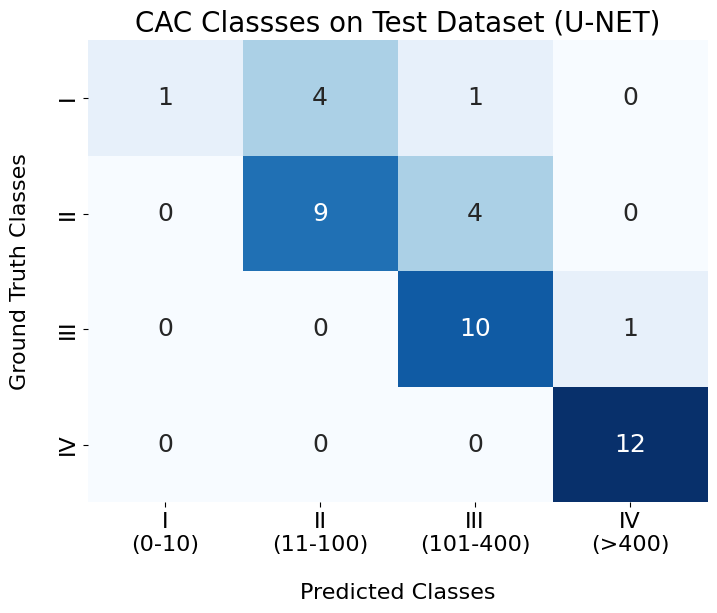}  
        \caption{Confusion Matrix for CAC class prediction using U-NET on the Test Dataset }
        \label{fig:unet_cac_scoring}
    \end{subfigure}
    \hfill
    \begin{subfigure}{0.45\textwidth}
        \centering
        \includegraphics[width=\linewidth]{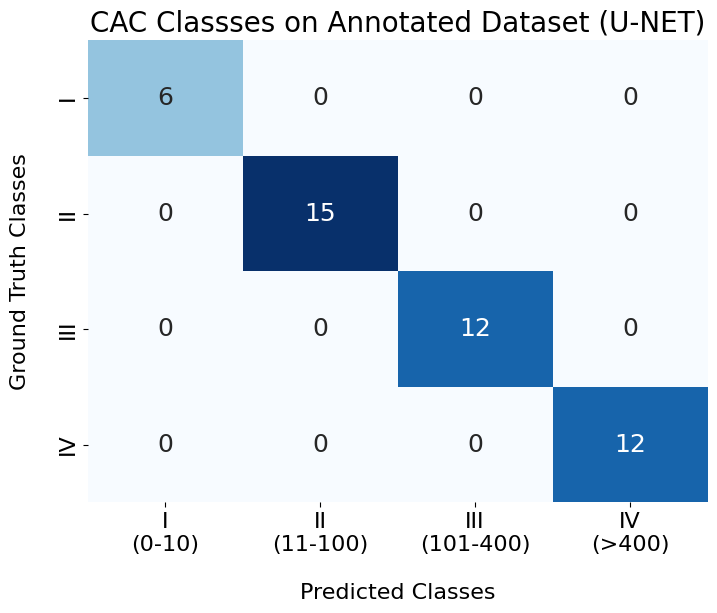}  
        \caption{Confusion matrix for CAC class prediction using U-NET on the annotated dataset, highlighting classification accuracy with annotated slices only.}
        \label{fig:unet_cac_scoring_gt}
    \end{subfigure}
    \caption{ CAC class predictions for each patient in the test dataset using U-NET.} 
    \label{fig:cac_comparison_unet_gt}
\end{figure*}

The CAC scoring results produced by the U-NET architecture are shown in Figure \ref{fig:unet_cac_scoring}, while Figure \ref{fig:unet_cac_scoring_gt} illustrates the CAC classification results using the annotated dataset. The discrepancies between these results indicate that CT slices without calcified areas have a negative impact on the overall classification performance.

\subsection{Performance Results of the Integrated System} 
The integrated system, combining DINO-LG based feature extraction, classification and U-NET segmentation, demonstrates significant improvement in identifying and scoring coronary artery calcification. In the classification part, the DINO-LG model efficiently distinguishes CT slices whether having calcified areas or not, ensuring that only relevant images are forwarded to the segmentation model. Identifying CT slices containing calcified areas has a crucial role to leverage segmentation task as it represented in Figure \ref{fig:cac_comparison_unet_gt}. \\ 

\begin{table*}[h!]
\caption{Comparison of classification performance metrics between UNET and Our Approach models across RCA, LAD, LCA, and LCX classes.}
\centering

\resizebox{1.0\textwidth}{!}{%
\begin{tabular}{lccc|ccc}
\toprule
\textbf{Metric} & \multicolumn{3}{c}{\textbf{UNET}} & \multicolumn{3}{c}{\textbf{Our Approach}} \\
\cmidrule(lr){2-4} \cmidrule(lr){5-7}
 & \textbf{Precision} & \textbf{Recall} & \textbf{F1-score} & \textbf{Precision} & \textbf{Recall} & \textbf{F1-score} \\
\midrule
RCA             & 1.00 & 1.00 & 1.00 & 1.00 & 1.00 & 1.00 \\
LAD             & 0.68 & 0.86 & 0.76 & \textbf{0.83} & 0.86 & \textbf{0.85} \\
LCA             & 0.62 & 0.59 & 0.60 & \textbf{0.71} & 0.59 & \textbf{0.64} \\
LCX             & 0.70 & 0.79 & 0.75 & \textbf{0.81} & 0.79 & \textbf{0.80} \\
\midrule 
\textbf{Avg. Accuracy}      & 0.75 & 0.81 & 0.78 & \textbf{0.84} & 0.81 & \textbf{0.82} \\ 
\bottomrule
\end{tabular}}
\vspace{0.3cm}
\label{tab:comparison_our_unet}
\end{table*}
The segmentation model used in the system is U-NET, chosen for its superior performance in IoU and Dice scores. A comparative analysis of the segmentation performance on the test dataset is conducted between the standalone U-NET and the integrated U-NET used within the system. The results, presented in Table \ref{tab:comparison_our_unet}, detail performance across four distinct classes: RCA, LAD, LCA, and LCX.

For the RCA class, both methods achieve IoU and Dice scores of 1.00. However, our approach excels particularly in the LAD and LCX categories, with a notable 0.85 F1-score in LAD and 0.80 in LCX, compared to the lower scores achieved by the standalone U-NET. The average accuracy of the integrated approach is also higher at 0.84 compared to U-NET’s 0.75, suggesting that the additional feature extraction via DINO-LG plays a significant role in enhancing classification reliability. The results show improved performance of the integrated system in LAD and LCX regions, where smaller calcifications are more prevalent.

\begin{figure*}[h!]
    \centering
    \begin{subfigure}{0.45\textwidth} % Adjusted width for better spacing
        \centering
        \includegraphics[width=\linewidth]{cac_class_conf_matrix.png}  
        \caption{Confusion Matrix for CAC Prediction using standalone U-NET}
        \label{fig:unet_cac_scoring_result}
    \end{subfigure}
    \hfill
    \begin{subfigure}{0.45\textwidth} % Adjusted width for better spacing
        \centering
        \includegraphics[width=\linewidth]{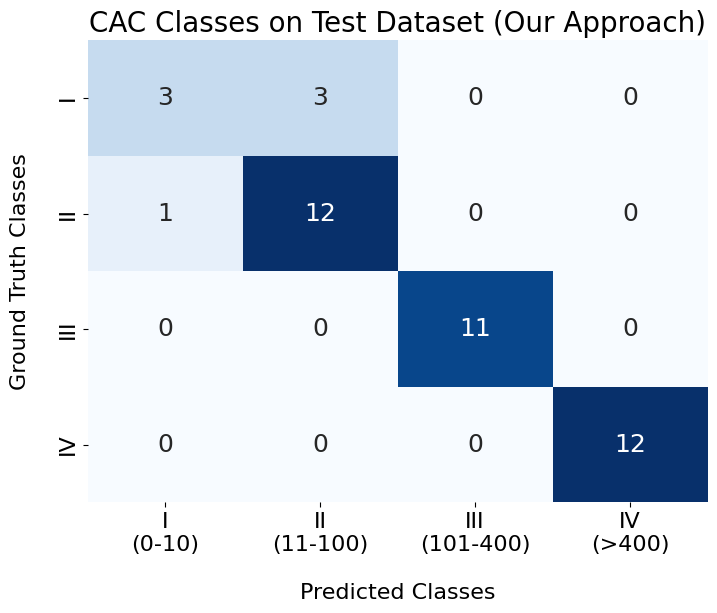}  
        \caption{Confusion Matrix for CAC Prediction using our proposed system}
        \label{fig:unet_cac_scoring_gt_result}
    \end{subfigure}
    \caption{Confusion matrices showing improved CAC classification accuracy of our proposed DINO-LG integrated system (b) compared to the standalone U-NET model (a) on the test dataset.} 
    \label{fig:cac_comparison_unet}
\end{figure*}
 
\begin{table*}[h!]
\fontsize{11pt}{11pt}\selectfont
\caption{Comparison of classification performance metrics for U-NET and Our System on CAC scoring.}
\centering
\begin{tabular}{lcc} 
\textbf{Metric}                     & \textbf{U-NET} & \textbf{Our Approach} \\ \hline
\vspace{0.03cm} \\ 
\multicolumn{3}{l}{\textbf{Low Risk Metrics (0-10)}} \\ \hline
Precision                           & 0.67                  & 1.00             \\  
Recall                              & 0.91                  & 1.00             \\  
F1-score                           & 0.77                  & 1.00             \\  
Sensitivity (TPR)                   & 0.17                  & 0.50             \\  
Specificity (TNR)                   & 1.00                  & 0.97             \\  
\vspace{0.01cm} \\ 
\multicolumn{3}{l}{\textbf{Moderate Risk Metrics (11-100)}} \\ \hline
Precision                           & 1.00                  & 0.75             \\  
Recall                              & 0.17                  & 0.50             \\  
F1-score                           & 0.29                  & 0.60             \\  
Sensitivity (TPR)                   & 0.69                  & 0.92             \\  
Specificity (TNR)                   & 0.86                  & 0.90             \\  
\vspace{0.01cm} \\ 
\multicolumn{3}{l}{\textbf{High Risk Metrics (101-400)}} \\ \hline
Precision                           & 0.69                  & 0.80             \\  
Recall                              & 0.69                  & 0.92             \\  
F1-score                           & 0.69                  & 0.86             \\  
Sensitivity (TPR)                   & 0.91                  & 1.00             \\  
Specificity (TNR)                   & 0.84                  & 1.00             \\  
\vspace{0.01cm} \\ 
\multicolumn{3}{l}{\textbf{Very High Risk Metrics ($>$ 400)}} \\ \hline
Precision                           & 0.92                  & 1.00             \\  
Recall                              & 1.00                  & 1.00             \\  
F1-score                           & 0.96                  & 1.00             \\  
Sensitivity (TPR)                   & 1.00                  & 1.00             \\  
Specificity (TNR)                   & 0.97                  & 1.00             \\  
\vspace{0.01cm} \\ 
\multicolumn{3}{l}{\textbf{Overall Metrics}} \\ \hline
Accuracy                            & 0.76                  & \textbf{0.90}           \\  
Weighted Avg Precision              & 0.80                  & \textbf{0.90}             \\  
Weighted Avg Recall                 & 0.76                  & \textbf{0.90}           \\  
Weighted Avg F1-score               & 0.73                  & \textbf{0.90}             \\  
Average Sensitivity (TPR)           & 0.69                  & \textbf{0.86}             \\  
Average Specificity (TNR)           & 0.92                  & \textbf{0.97}             \\ 
\end{tabular}
\vspace{0.3cm}
\label{tab:comparison_cac_scores}
\end{table*}

The segmentation results generated by standalone U-NET and our system are also utilized for calculating CAC scores which are subsequently categorized into risk classes based on these scores. As it is illustrated in Figure \ref{fig:cac_comparison_unet}, our proposed system demonstrates a more refined classification performance compared to the standalone U-NET, especially in distinguishing between moderate and high risk classes.

The comparative results presented in Table~\ref{tab:comparison_cac_scores}  clearly demonstrate the superiority of our integrated system over the standalone U-NET model in coronary artery calcium (CAC) scoring. Our integrated system achieves consistent improvements across all risk categories, particularly excelling in precision, recall, and F1-scores for both low-risk and high-risk categories. For instance, in the "Low Risk" category, precision, recall, and F1-score all reach perfect values of 1.00 in the proposed system, whereas the U-NET model struggles with a precision of 0.67 and F1-score of 0.77. 

Sensitivity and specificity metrics further underline the system's effectiveness. Average sensitivity (TPR) across all risk categories rises from 0.69 in the U-NET model to 0.86 in the proposed system, demonstrating improved detection of calcified areas. Similarly, specificity (TNR) increases from 0.92 to 0.97, reducing false positives and enhancing the model's reliability.

Moreover, the overall weighted averages for precision, recall, and F1-score in our approach are 0.90 each, substantially outperforming the corresponding values of 0.80, 0.76, and 0.73 in the U-NET model. These advancements signify a robust and efficient framework, ensuring accurate classification and segmentation of calcifications while minimizing false negatives and false positives. This enhanced performance is particularly beneficial in clinical settings, where accurate CAC assessment directly impacts patient management and treatment outcomes.

\section{Discussion and Conclusion} 

This study introduces DINO-LG, a label-guided self-supervised learning approach for coronary artery calcium detection. Our key findings demonstrate that incorporating targeted guidance during DINO training improves feature discrimination for small, sparse calcifications. Quantitatively, DINO-LG achieved 89\% sensitivity and 90\% specificity for calcified slice detection, representing 10-13 percentage point improvements over standard DINO. When integrated with U-NET segmentation, the complete system achieved 90\% accuracy in CAC risk classification compared to 76\% for standalone segmentation.

Statistical analysis confirms the significance of these improvements. McNemar's test on classification results yielded p < 0.001, demonstrating that the performance improvement is statistically significant. The classification stage reduced false-negative and false-positive rates by 49\% and 57\% respectively, minimizing both missed diagnoses and unnecessary computational processing.

The U-NET architecture achieves the highest average performance across all coronary artery classes with IoU scores of 0.97, 0.61, 0.43, and 0.49 for RCA, LAD, LCA, and LCX respectively (Table \ref{tab:segmentation_scores}). The variation in performance across different arteries reflects the inherent challenges of detecting smaller calcifications in LAD and LCX compared to the typically larger and higher-contrast calcifications in RCA. Our integrated system partially addresses this limitation, achieving improved F1-scores of 0.85 for LAD and 0.80 for LCX compared to 0.76 and 0.75 for standalone U-NET.

The training method we propose for DINO is not limited to coronary artery calcium detection. This approach can be applied to various medical imaging tasks with sparse annotations, including lung nodule detection, liver lesion identification, and tumor segmentation. The ability of vision foundational models to capture multiple feature patterns simultaneously suggests potential for cross-domain learning, where training across multiple anatomical regions could enhance the model's ability to learn diverse and transferable representations.

\textbf{Comparison with Prior Work.} Compared to recent deep learning approaches for CAC scoring \cite{Eng2021-wv, Zeleznik2021-ki}, our method offers several advantages. While Eng et al. \cite{Eng2021-wv} achieved 71-94\% sensitivity using supervised learning on dedicated datasets, our self-supervised approach requires fewer annotated samples while maintaining competitive performance (89\% sensitivity). The classification-then-segmentation pipeline also provides computational advantages by processing only relevant slices, reducing inference time by approximately 60\% compared to volume-wide segmentation approaches.

Our unified framework handles both ECG-gated and non-gated CT acquisitions, addressing a practical clinical need that most prior work has not explored systematically. This versatility is particularly valuable since non-gated scans are more commonly performed in routine clinical practice, yet CAC scoring has traditionally relied on dedicated gated acquisitions.

\textbf{Interpretation of Results.} The superior performance of DINO-LG stems from its ability to learn calcification-specific features through targeted augmentation. Analysis of the attention maps (Figure \ref{fig:attention_maps}) reveals that label-guided training directs the model's self-attention mechanisms toward high-density regions corresponding to calcifications. This targeted attention is particularly beneficial for detecting small calcifications in LAD and LCX arteries, where standard approaches often fail due to size and contrast limitations.

The PCA visualization demonstrates that label-guided training produces features with concentrated activation on calcified regions, contrasting sharply with the diffuse patterns observed in standard DINO features. This suggests that the guidance signal successfully steers the self-supervised learning process toward task-relevant features without compromising the model's ability to learn general representations from unlabeled data.

\textbf{Clinical Implications.} The high specificity (90\%) of our classification stage reduces unnecessary segmentation processing, which is important for clinical workflow integration. However, the 11\% false-negative rate in slice classification remains a concern for clinical deployment, as missed calcifications could lead to underestimated risk scores. Future work should focus on further reducing this rate through ensemble methods or uncertainty quantification.

The 90\% accuracy in risk category classification demonstrates potential clinical utility, though validation on larger multi-center datasets with clinical outcome correlation is needed to establish the system's impact on patient management decisions. The ability to process both gated and non-gated scans with a single model could facilitate opportunistic CAC screening from routine chest CT examinations, potentially identifying at-risk patients who would not otherwise undergo dedicated cardiac imaging.

Despite promising results, several limitations should be acknowledged. The scale of our training dataset (1,003 patients total) is relatively small for foundational model development, which typically requires millions of diverse images. This limits generalizability to different populations, imaging protocols, and scanner types. Our evaluation is based on a single institutional dataset, lacking multi-center validation that would better establish robustness across diverse clinical settings. The study focuses on technical performance metrics without correlation to long-term cardiovascular outcomes, which is essential for demonstrating clinical utility. Detection of very small calcifications (< 1mm) remains challenging, particularly in cases with high image noise, and the impact of inter-observer variability in ground truth annotations on model stability has not been thoroughly investigated.

Additional limitations not previously mentioned include: (1) The U-NET segmentation performance varies significantly across coronary arteries (Dice: 0.98 for RCA vs 0.60 for LCA), suggesting the need for artery-specific models or attention mechanisms. (2) Our evaluation lacks temporal consistency analysis, we did not assess whether the model produces stable predictions for the same patient scanned at different time points. (3) The computational overhead of the three-stage pipeline (feature extraction, classification, segmentation) may limit real-time application in high-throughput clinical environments, despite the reduction in processed slices. (4) We did not investigate the model's robustness to common CT artifacts such as motion blur, metal artifacts from stents, or beam-hardening effects, which are frequently encountered in clinical practice.

Future work should address these limitations through multi-center validation studies, larger-scale pre-training on diverse medical imaging datasets, integration of clinical outcome measures, development of artery-specific segmentation approaches, investigation of temporal consistency for longitudinal tracking, and systematic evaluation of robustness to common CT artifacts and image quality variations.

In conclusion, this work demonstrates that label-guided self-supervised learning can effectively direct Vision Transformers to learn task-specific features while maintaining the benefits of pre-training on unlabeled data. DINO-LG achieves substantial improvements in coronary artery calcium detection and scoring, offering a promising approach for medical imaging tasks where labeled data is scarce and regions of interest are small and sparse. The integration of self-supervised learning, targeted classification, and segmentation provides both improved accuracy and computational efficiency, with potential applications extending beyond CAC detection to other medical imaging domains facing similar challenges.

\textbf{Acknowledgements} The project described was supported by the NIH National Center for Advancing Translational Sciences through grant number UL1TR001998. The content is solely the responsibility of the authors and does not necessarily represent the official views of the NIH. We thank the Scientific and Technological Research Council of Turkiye (TUBITAK) for supporting the author Caner Ozcan through BIDEB-2219 International Postdoctoral Research Fellowship Program (grant no. 1059B192300232). \\
\textbf{Conflict of interest} The authors declare no competing interests.

\bibliography{sn-bibliography}

\end{document}